\documentclass{iaus}
\usepackage{graphicx}

\title[Observations of pre-stellar cores] 
{Observations of pre-stellar cores}

\author[M. Tafalla]   
{M. Tafalla}%

\affiliation{Observatorio Astron\'omico Nacional, Alfonso XII 3, E-28014 Madrid,
Spain \break email: m.tafalla@oan.es}

\pubyear{2005}
\volume{231}  
\pagerange{xxx--xxx}
\date{?? and in revised form ??}
\jname{Astrochemistry: Recent Succeses and Current Challenges}
\editors{Dariusz C. Lis, Geoffrey A. Blake \& Eric Herbst, eds.}
\begin{document}

\maketitle

\begin{abstract}
Our understanding of the physical and chemical structure of pre-stellar cores,
the simplest star-forming sites,
has significantly improved since the last IAU Symposium on Astrochemistry 
(South Korea, 1999). Research done over these years has revealed 
that major molecular species like CO and CS systematically 
deplete onto dust grains at
the interior of pre-stellar cores, while species like N$_2$H$^+$ and
NH$_3$ survive in the gas phase and can usually be detected 
towards the core centers. Such a selective behaviour 
of molecular species gives rise to
a differentiated (onion-like) chemical composition,
and manifests itself in molecular maps as a dichotomy between
centrally peaked and ring-shaped distributions. From the point of view
of star-formation studies, the identification of molecular
inhomogeneities in cores helps to resolve
past discrepancies between observations made using different 
tracers, and brings the possibility of self-consistent modelling
of the core internal structure. Here I present recent work on 
determining the physical and chemical structure of two 
pre-stellar cores, L1498 and L1517B, using observations
in a large number of molecules and Monte Carlo radiative
transfer analysis. These two cores are typical examples of 
the pre-stellar core population, and their chemical composition
is characterized by the presence of large freeze out holes in most 
molecular species. In contrast with these chemically processed objects,
a new population
of chemically young cores has started to emerge. The characteristics
of its most extreme representative, L1521E, are briefly reviewed.
\keywords{astrochemistry, molecular processes, radiative transfer, stars: formation,
ISM: clouds, ISM: molecules, radio lines: ISM}
\end{abstract}

\firstsection 
\section{Pre-stellar cores as star-forming sites}

Pre-stellar (or starless) cores are the simplest star-forming 
sites. They are 
isolated, lie nearby, and form one star (or one binary) at the time;
they closely resemble the theorist's ideal of a star forming region.
When observed in a molecular tracer like ammonia, a pre-stellar 
core appears as a centrally concentrated structure containing 
one or a few solar masses of material and having a typical size 
of about 0.1 pc (see Figure 1 and Benson \& Myers 1989 for further global
properties of cores). 

\begin{figure}
\begin{center}
\includegraphics{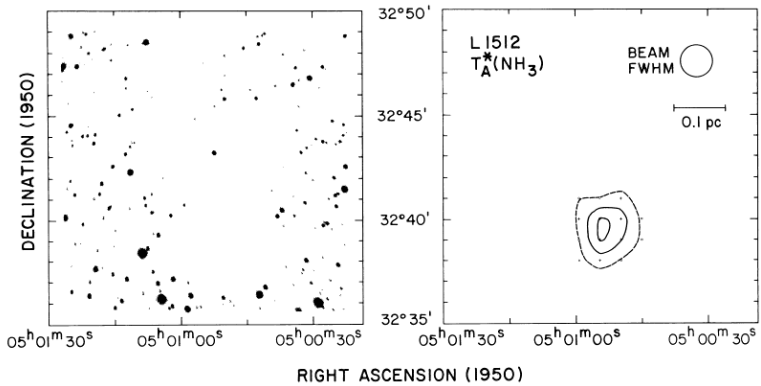}
   \caption{A representative pre-stellar 
   core, L1512. Left: optical image from the
Palomar Sky Survey, where L1512 appears as a patch of obscuration against
the background of stars. Right: map of the NH$_3$ emission toward the same
region showing L1512 as a centrally concentrated (slightly resolved)
object. Figure from Benson \& Myers (1989).
}\label{fig:fig1}
\end{center}
\end{figure}

Pre-stellar cores are the dominant star-forming sites in nearby
molecular clouds like Taurus-Auriga, where stars like our Sun are 
currently forming in the so-called
``isolated mode'' (e.g., Shu, Adams, \& Lizano 1987). They are
not, however, the dominant star-forming regions of our galaxy,
as most stars in the Milky Way have formed in groups
(e.g., Adams \& Myers 2001), and therefore must
result from the collapse of more complex gas structures.
Still, star formation in isolated pre-stellar
cores seems to involve most of the physical elements that we 
associate with the birth of a typical low-mass star, like gravitational
infall, disk formation, and bipolar outflow ejection. All 
these elements, in fact, were first identified in stars forming 
in isolated cores, and they can be studied with great detail in these
simple environments.

The above reasons of simplicity make pre-stellar cores ideal sites
to study the initial conditions of star-formation (e.g., Ward-Thompson 
et al. 1999). Cores are also the most promising places to test the different
competing models of star formation, in particular the (fast) turbulence
driven scenario (Mac Low \& Klessen 2004) and the (slow) magnetically-mediated 
star formation models (Shu et al. 1987, Mouschovias \& Ciolek 1999). By determining how
pre-stellar cores contract out of the more diffuse ambient cloud and
by what mechanism cores lose their gravitational
support and start collapsing to form a star, we may be able to
observationally distinguish between these two models.

\section{Cores as laboratories of ISM chemistry}

Pre-stellar cores are also some of the simplest chemical laboratories
in the interstellar medium (ISM) because of their isolation,
simple (close to spherical) geometry, and low (10~K), close to
constant gas temperature. Despite this simplicity, however, 
pre-stellar cores are not chemically homogeneous, and 
the ignorance of this fact
has caused in the past serious difficulties when attempting to
derive the core internal structure from observations. Early
warnings of chemical inhomogeneities were the striking
discrepancies found when mapping cores using different molecular
tracers. A classical example of this problem are the observations
made using NH$_3$ and CS, two molecular species expected to trace similar
gas conditions. As found by Zhou et al. (1989), maps
made in NH$_3$ and CS present systematically different sizes
(NH$_3$ maps are a factor of 2 smaller), different shapes, and
often different peak positions. Resolution effects or
radiative transfer complexities were soon found insufficient to
explain these discrepancies.

A hint of a solution to the tracer discrepancies
comes from the recent realization that species like CO and CS are 
systematically depleted at the
centers of cores due to freeze out onto cold dust grains 
(Kuiper et al. 1996, Willacy et al. 1998, Kramer et al. 1999, 
Caselli et al. 1999). The identification of molecular freeze out 
in cores
(long expected from theoretical grounds, see Watson \& Salpeter 1972)
has renewed the interest in the study of dense core chemistry, and
it brings promise of resolving the old tracer discrepancies
with relatively simple chemical processes. From the point of view of
star-formation studies, the identification of freeze out in some molecules
is forcing a re-evaluation of the behaviour
of the different dense gas tracers under typical core conditions, as we
depend on them to infer basic core properties like gas temperature and
kinematics. Determining the chemical structure of cores has therefore 
become a 
necessary step in our attempt to understand how stars are born.

\section{A molecular survey of L1498 and L1517B}

To understand the chemical behaviour of dense gas tracers
in star-forming regions, we have carried out a systematic 
molecular survey of two pre-stellar cores in
the Taurus-Auriga cloud complex, L1498 and L1517B (Tafalla,
Myers, Caselli, \& Walmsley 2004; Tafalla et al. 2005 in preparation).
We selected these two cores (Fig. 2) for being isolated, close to
round, and otherwise typical cores of their surrounding cloud, and we
have observed them in the 1.2 mm continuum and in a large number of 
molecular lines from 13 different species using the IRAM 30m, Effelsberg 
100m, and FCRAO 14m radio telescopes.

\begin{figure}
\begin{center}
\includegraphics[height=2in]{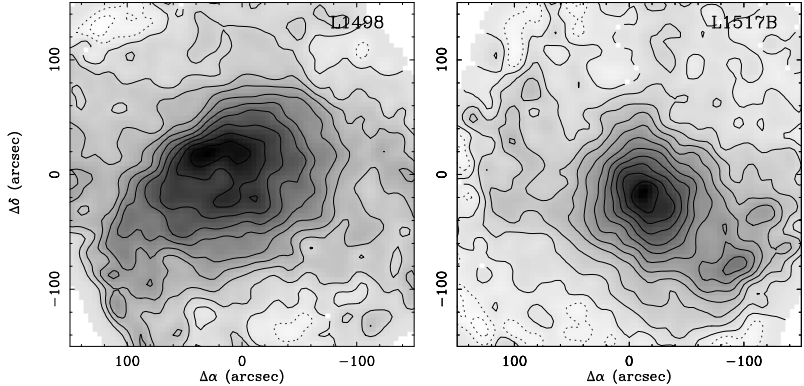}
   \caption{1.2mm continuum maps of L1498 (left) and L1517B (right). Note the
   central concentration and close-to-round shape of the emission. First
   contour and contour spacing are 2 mJy/$11''$-beam.
}\label{fig:fig2}
\end{center}
\end{figure}

The goal of this project is to model self-consistently all the 
observed emission
in order to determine how the different species trace the core interior,
and to provide a high quality set of molecular abundances for testing
chemical models. We can divide the analysis of the observations in
two steps. First, we determine the physical parameters of the 
cores by modelling their distributions of density, temperature, and
gas kinematics assuming that the cores are spherically symmetric. 
Once these parameters have been fixed, 
the cores can be seen as laboratories of known physical properties, and
the abundances of the different molecular species can be derived
directly by fitting their observed emission.

\subsection{Physical structure of L1498 and L1517B}

To derive the density profiles of L1498 and L1517B we rely on the 
dust continuum emission, as this is the most unbiassed tracer
of the core column density. We assume a
1.2mm dust emissivity of 0.005 cm$^2$ g$^{-1}$ and a dust temperature
of 10~K, and we note that these parameters are
the largest source of uncertainty of the whole
analysis (see Tafalla et al. 2004 for further details). By fitting
the radial profiles of 1.2mm continuum emission, we find density profiles
very close to those expected for isothermal spheres with
central densities of $10^5$ and $2\times 10^5$ cm$^{-3}$
for L1498 and L1517B, respectively (see Alves et al. 2001 
and Evans et al. 2001 for similar fits to other pre-stellar cores).

\begin{figure}
\begin{center}
\includegraphics[height=4.5in]{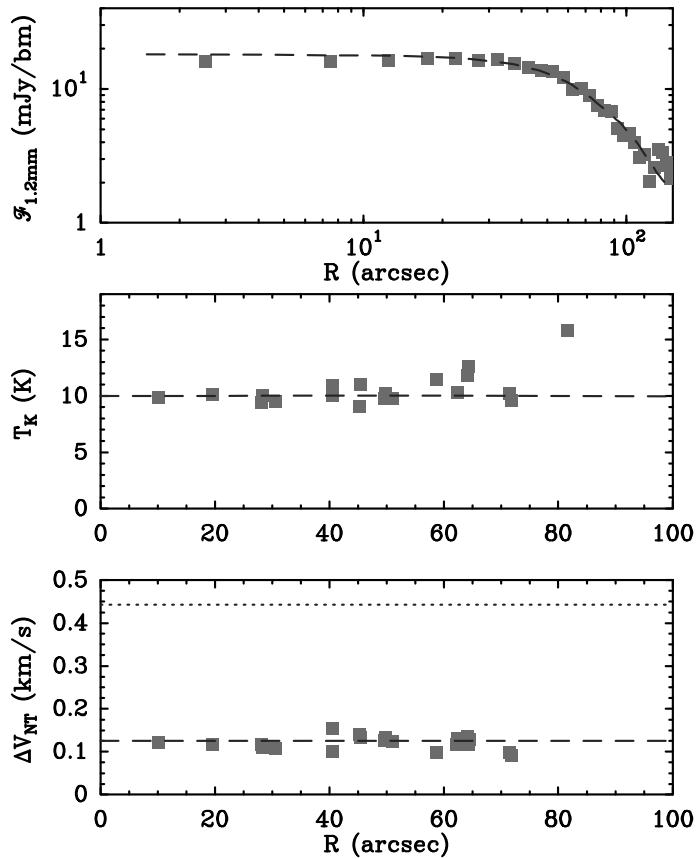}
   \caption{Determination of the physical parameters of L1498. Top: radial
   profile of 1.2mm continuum emission from the map of Fig. 2 
   (squares) and prediction from the best fit density
   determination (dashed lines). Middle: NH$_3$-derived gas kinetic
   temperature estimate (squares) and constant 10~K fit (dashed lines).
   Bottom: non thermal linewidth component derived from NH$_3$(1,1) 
   spectra (squares) and constant component with FWHM = 0.125 km/s (dashed lines).
   The dotted line indicates the expected value for a sonic component.
   }\label{fig:fig3}
\end{center}
\end{figure}

To derive the gas temperature profile, we use the NH$_3$ emission, 
that we will see below traces well the inner core. From the combined 
analysis of the emission from the meta-stable J,K=1,1 and 2,2 levels,
we derive constant temperature profiles for both cores with values 
close to 10~K. 

Finally, we derive core turbulent profiles using the linewidth of
NH$_3$(1,1) complemented with other species. We subtract the
(constant) thermal component and find an also constant turbulent 
component of less than 1/3 the speed of sound. Such a low
level of turbulence seems already problematic for current turbulent
models of star formation.

\subsection{Chemical structure of L1498 and L1517B}

Once the physical structure of each core has been determined, the only free
parameter left to fit the observed emission of a given species is the radial
distribution of its abundance. To convert this distribution into a 
predicted line intensity, we use a Monte Carlo radiative transfer code
that assumes spherical symmetry (Bernes 1979). For each observed transition,
we require that the model fits both the radial distribution of integrated
intensity (derived by averaging azimuthally the data) and the
line spectrum observed toward the core center. When possible, we use
the thin emission of a rare isotopomer (like C$^{18}$O and C$^{34}$S)
to determine the abundance of the major species (like CO and CS).
As for most molecules we have observed two or
more transitions, their abundance profile is over-determined by the data. 

In a meeting like this one is important to emphasize that the 
analysis presented here 
can only be carried out if certain molecular parameters have been
previously determined. Among these parameters, the collision rates 
are critical because they
regulate the excitation of the levels and therefore the relation between
model abundance and predicted intensity. In fact, the availability
of collision rates with H$_2$ or He was a main criterion for selecting
the species observed in this survey. Another important molecular
parameter needed to fit the narrow lines observed in L1498 and L1517B is
the frequency of each transition. For our modelling, accuracies of 10 kHz are
required, and although this level
is now easily achievable in laboratory work, not all important molecular
transitions have yet been measured with this detail; accurate laboratory 
frequencies of molecular
ions are in particular absent. For our work, we have used the most recent
frequency determinations from the CDMS, Gottlieb and collaborators,
Dore and collaborators, and the JPL catalog.

\begin{figure}
\begin{center}
\includegraphics[width=3.5in,angle=270]{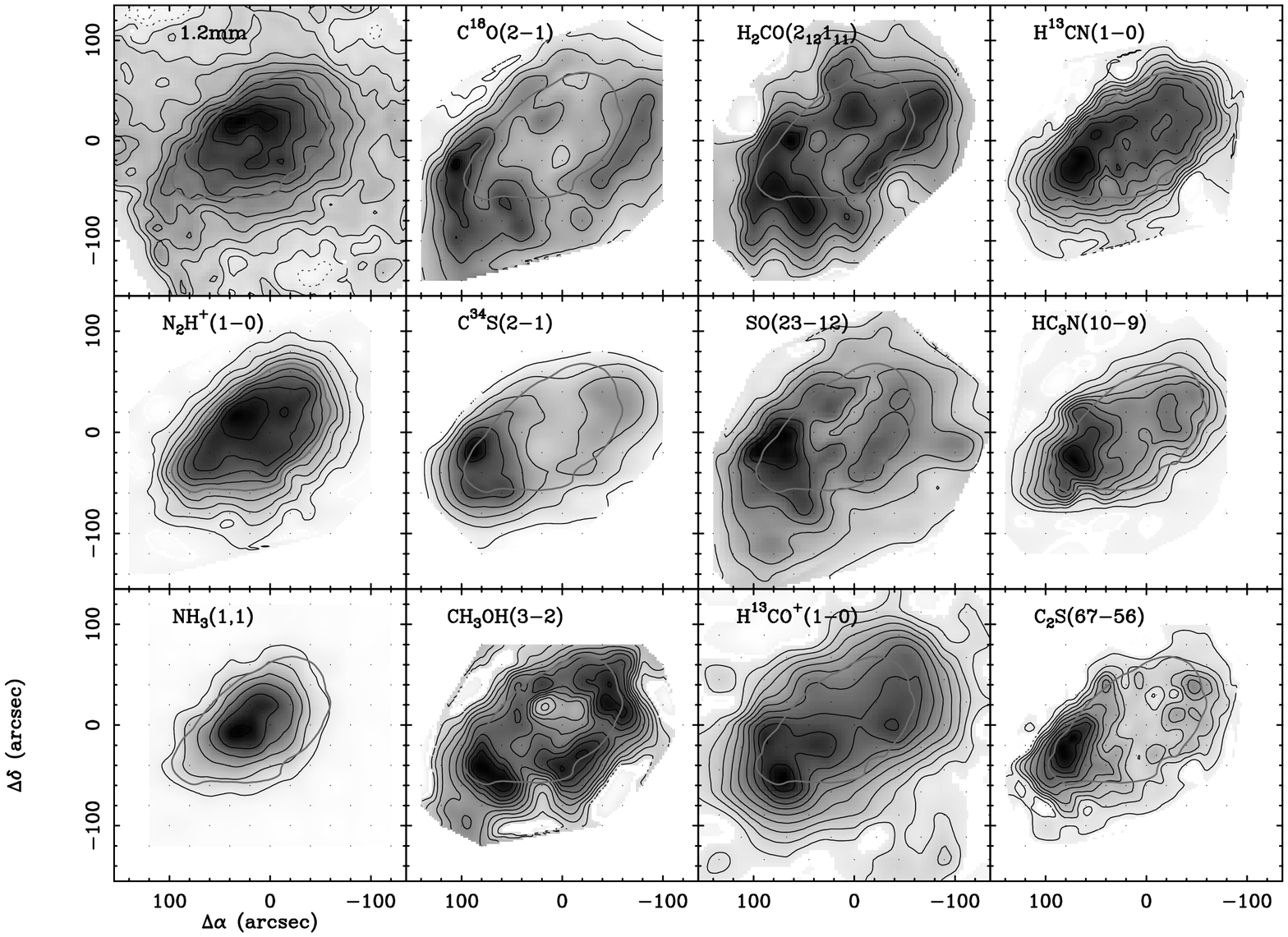}
\includegraphics[width=3.5in,angle=270]{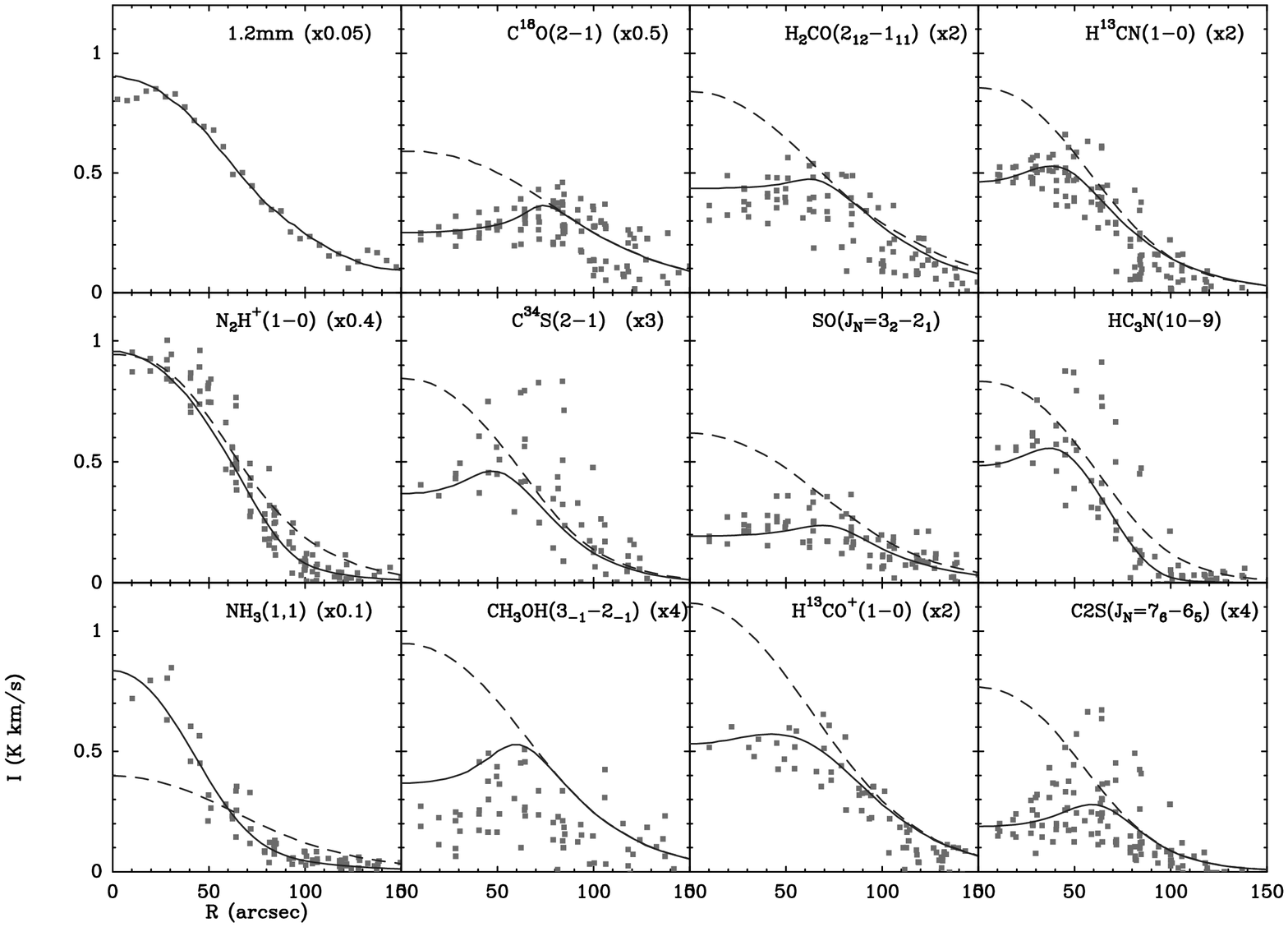}
   \caption{Partial results of the molecular survey toward L1498.
   Top: maps of 1.2mm continuum and molecular lines for a 
   sample of transitions. The three
   panels in the leftmost column contain centrally concentrated emission maps, 
   while the rest of the panels show ring-shaped distributions. Bottom: Radial 
   profiles of emission derived from the maps in the top (squares) together
   with the emission predictions from two Monte Carlo models. The dashed
   lines are the predictions from constant abundance models set to fit the
   outer core emission, while the solid lines are predictions from the
   best-fit models.
      }\label{fig:fig4}
\end{center}
\end{figure}

To illustrate the process of deriving abundance profiles from the
data, we present in Fig. 4 (top) a sample of integrated intensity maps for
the L1498 core. The three panels in the left column 
show centrally concentrated
distributions corresponding to the 1.2mm continuum, N$_2$H$^+$(1--0),
and NH$_3$(1,1). The rest of the maps, on the other hand, 
show ring-like distributions
with a relative minimum at the dust peak. These
ring-like distributions are very fragmented and slightly different 
for each molecule, although there are systematic features, like a brighter
peak to the southeast. 

When we convert the above maps into radial profiles of integrated 
intensity and attempt to fit them with different 
abundance profiles, we obtain the results shown
in the bottom part of Fig. 4. For each molecule, this figure
shows two model predictions: a
constant abundance model chosen to fit the emission in the outer core
(dashed lines) and a best-fit model (solid lines).
As the figure shows, the N$_2$H$^+$ emission is fit reasonably well by the
constant abundance model, while the observed NH$_3$ emission is more centrally
concentrated than predicted by a constant abundance model. This 
species, therefore, requires a significant enhancement of abundance
toward the core center.

In agreement with the expectation from the ring-shaped maps, the
rest of the molecules cannot be fit with a constant abundance model,
as these models clearly overestimate the central intensity by a
factor of 2 or more when forced to fit the outer core emission.
Only using a sharp central abundance drop can both the outer and
inner emission be simultaneously fit. For this reason, we
have chosen for our best fit models simple step functions with
close to zero abundance toward the center. From the quality
of the fit, we conclude that the data are consistent with 
a (close to) total absence of certain molecules at the core center.

Although not shown here for lack of space, the abundance results for
L1517B are very similar to those for L1498 (a full account of the
analysis will be presented in Tafalla et al. 2005, in preparation).
The outer abundances for both cores are in fact rather close, and 
for most species they agree within a factor of 2; this suggests that 
both objects have contracted from gas having similar chemical
compositions. The size of the central hole, on the other hand, 
is different in the two cores. L1517B presents significantly 
smaller central molecular holes (about 50\% smaller than L1498), which may be
related to the more concentrated gas distribution found in this core.

\subsection{Consequences for dense core studies}

L1498 and L1517B seem in every aspect typical pre-stellar cores.
More restricted studies of other systems by different authors,
as those presented in this meeting by the posters of Butner et al., 
Buckle et al., Friesen et al., Zinchenko et al., show patterns of molecular
abundance very similar to those found in L1498 and L1517B; it seems
therefore
natural to assume that the abundance profiles of L1498 and
L1517B are representative of the 
pre-stellar core population
as a whole (see also below for exceptions). Thus, 
despite their expected simplicity, 
pre-stellar cores must have a strongly differentiated chemical
composition. 

If chemical inhomogeneities are part of the initial
conditions of star formation, they need to be considered 
seriously when sampling star-forming gas with molecular tracers.
The bottom panels in Fig. 4, in particular, show that many
molecular maps of a core may reflect more its chemical
composition than its physical structure, and illustrate
the danger in attributing emission peaks of even thin lines
like C$^{34}$S(2--1) to real core sub-structure. A main
lesson from this analysis is therefore that one needs to be 
careful when choosing dense gas tracers, and that N$_2$H$^+$
and NH$_3$ are some of the best choices we have. These species,
together with their isotopomers and the H$_2$D$^+$ ion
recently detected by Caselli et al. (2003) in the pre-stellar
core L1544, are probably all
the molecular tracers left to study the conditions in the inner
core (even N$_2$H$^+$ may deplete at high densities, see 
Bergin et al. 2002 and Pagani et al. 2005). Posters in this meeting by Crapsi et al. 
and Vastel et al. show the use of these more reliable tracers 
to study the central conditions of pre-stellar cores.

The modelling of pre-stellar core chemistry is the subject of
several contributions in this meeting (talks by Shematovich and Roberts,
and posters by
Aikawa et al., Rawlings, Lee et al., Walmsley et al.), 
so I refer to them for further details on this topic. The only point
worth emphasizing here is that the unique behaviour of species
like N$_2$H$^+$ and NH$_3$ strongly suggests that their
survival in the gas phase results from a low binding energy of N$_2$,
as initially suggested by Bergin \& Langer (1997) and Charnley (1997).
However, similar binding energies for N$_2$ and CO have been 
recently measured by \"Oberg et al. (2005), a result that has been used
to claim that other mechanisms, like a low sticking coefficient for 
molecular or atomic N, may be needed
to explain the behaviour of N$_2$H$^+$ and NH$_3$ (Flower et al.
2005). Further work in this topic is clearly needed 
if we are to understand and use to our advantage 
the peculiar chemistry of these two important dense gas tracers.

\section{Searching for young pre-stellar cores}

If we observe pre-stellar cores selected from 
NH$_3$ surveys (like those from Myers and collaborators),
we find that
molecular depletion is the norm without exception. This 
systematic trend suggests that NH$_3$-selected cores are
significantly advanced in their process of gas contraction
from the ambient cloud, and that there must exist a population
of younger cores that are their precursors. Identifying such a 
population of cores is interesting not only because of their chemical
properties, but because its members should constitute a missing link
between cloud and core conditions, and they will therefore provide
useful clues on the (mysterious)
physical process that drives core condensation out of ambient material.

\begin{figure}
\begin{center}
\includegraphics{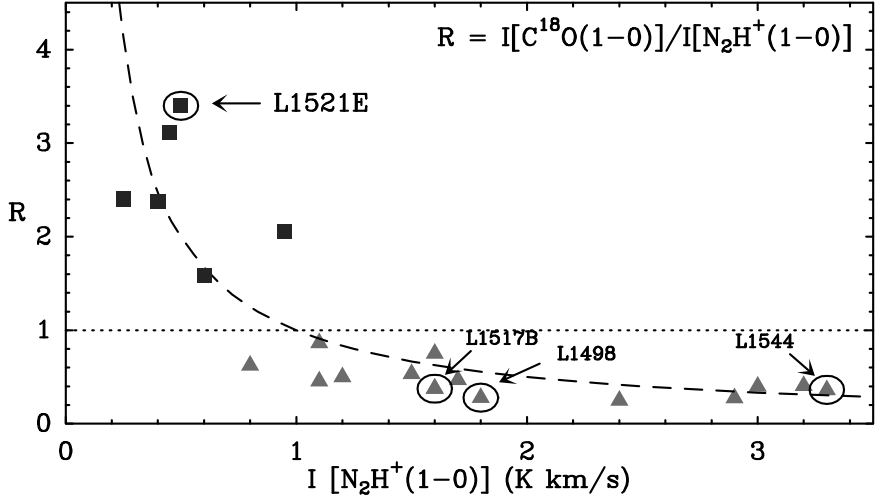}
   \caption{R (=I[C$^{18}$O(1-0)]/I[N$_2$H$^+$(1-0)]) as function of 
   I[N$_2$H$^+$(1-0)] for two samples of cores. The triangles are
   ``classical'' starless cores selected for their easily 
   detectable NH$_3$ emission,
   while the squares are starless cores selected for their weak 
   NH$_3$ emission. The dotted horizontal line marks the expected 
   approximate boundary between cores with and without C$^{18}$O freeze
   out, and the dashed line is the prediction from a toy model
   of core chemical evolution. Note how L1521E stands out among the young 
   core candidates (R$>1$).
}\label{fig:fig5}
\end{center}
\end{figure}

To search for young pre-stellar cores, we can take advantage of
their expected chemical properties. One of them is a
lower degree of CO freeze out at the center, which should make a
young pre-stellar core appear centrally concentrated in a C$^{18}$O
map. Another expected property is a low abundance of late-time
species like N$_2$H$^+$ and NH$_3$, which will translate into weak
emission in any line of these tracers. To relate these properties, 
we define the intensity ratio
$$R = {I[\mathrm{C}^{18}\mathrm{O}(1-0)] \over I[\mathrm{N}_2\mathrm{H}^+(1-0)]},$$
where the intensities are evaluated at the core center. This ratio
is an easily measurable quantity, as both lines lie in the 3mm
wavelength band and can therefore be observed with similar
angular resolution. Young cores are expected to have relatively large
central C$^{18}$O(1--0) emission together with weak N$_2$H$^+$(1--0)
lines, so they are expected to have relatively large values of the $R$ 
parameter. Conversely, old cores are expected to have relatively
weak central C$^{18}$O(1--0) emission and strong N$_2$H$^+$(1--0) lines,
so they should be characterized by 
relatively low values of $R$. The approximate boundary
between these two behaviours can be calculated using the ratio predicted 
by our Monte Carlo model for cores like L1498 and L1517B
assuming constant abundances for the two species. In this way, we find
the expected separation between young and old cores near 
the value $R=1$.

As mentioned before, searches for young cores using NH$_3$-selected
candidates seem doomed to fail, and this is illustrated by the
triangles in Fig. 5, which correspond to a survey of that type of
objects carried out with the FCRAO telescope. In this figure, 
all selected objects lie below the $R=1$ line,
including (not surprisingly) L1498 and L1517B studied before. The
objects from this survey probably span a range of ages, as
suggested by the range of N$_2$H$^+$(1--0) intensities that 
starts near 1 and ends past 3 with evolved cores like L1544
(e.g., Crapsi et al. 2005). However, they seem to be missing
the youngest cores.

To identify candidates to young starless cores we need to include 
cores having weak NH$_3$ emission, and this has been done selecting sources
from the survey of Suzuki et al. (1992) and from a survey of the
L1521 filament in Taurus using the FCRAO telescope. These cores 
finally fill the region in the plot expected for young cores, as they have
low N$_2$H$^+$(1--0) intensities together with large $R$ values.
Among these objects, L1521E has the largest $R$ ratio (=3.4), and
therefore appears as the best candidate for a young starless core.
Previous suggestions of this core being extremely young have been
made by Suzuki et al. (1992) and Hirota et al. (2002).


Given the unusual characteristics of L1521E, we have carried out
a molecular survey of this core in a similar manner as we have studied 
L1498 and L1517B. From a preliminary analysis of these data,
we conclude that it has no significant CO or CS central
depletion (Tafalla \& Santiago 2004), and that its N$_2$H$^+$ 
and NH$_3$ abundance is 8 times lower and 20 times lower than
L1498 and L1517B. These characteristics truly classify L1521E 
as a chemically young pre-stellar core, and therefore suggest that
this object has contracted from the ambient cloud 
to its observed state rather
recently. Surprisingly, however, L1521E has a central density
of $10^5$ cm$^{-3}$, which is very similar to that found in L1498 
and L1517B. This high density contradicts the expectation that a young
core should be less dense, and suggests that L1521E may have
contracted faster than the others (also Aikawa et al. 2005).
Clearly more work is needed to clarify the origin 
of this group of starless cores, and recent studies of similar
systems are encouraging (Hirota et al. 2004, Morata et al. 2005).
The poster by Hirota and Yamamoto in this meeting
provides recent results on this topic. 

\begin{acknowledgments}
I thank the organizers for their invitation
and for a highly enjoyable and productive
meeting. Part of the work presented here is the result of an
ongoing collaboration with Joaqu\'{\i}n Santiago, Phil Myers,
Paola Caselli, Malcolm Walmsley, Claudia Comito, and Antonio
Crapsi. I thank them for help and discussions on pre-stellar cores 
over the last several years.
\end{acknowledgments}

\end{document}